\documentclass[aps,prb,amsmath,twocolumn]{revtex4-1}
\usepackage{graphicx}

\begin{document}

\title{Dielectric response of Anderson and pseudogapped insulators}

\author{M.~V.~Feigel'man$^{1,2}$,  D.~A.~Ivanov$^{3,4}$ and E.~Cuevas$^{5}$}
\date{January 9, 2018}

\affiliation{$^{1}$ L.~D.~Landau Institute for Theoretical Physics, Chernogolovka, 142432,
Moscow region, Russia}
\affiliation{$^{2}$ Skolkovo Institute for Science and Technology, Moscow, Russia}
\affiliation{$^{3}$ Institute for Theoretical Physics, ETH Z\"urich, 8093 Z\"urich, Switzerland}
\affiliation{$^{4}$ Department of Physics, University of Z\"urich, 8057 Z\"urich, Switzerland}
\affiliation{$^{5}$ Departamento de F\'isica, Universidad de Murcia, E-30071,  Murcia, Spain}

\begin{abstract}
{Using a combination of analytic and numerical methods, we study the polarizability
of a (non-interacting) Anderson insulator in one, two, and three dimensions and demonstrate 
that, in a wide range of parameters,
it scales proportionally to the square of the localization length, contrary to earlier claims based
on the effective-medium approximation. We further analyze the effect of electron-electron interactions
on the dielectric constant in quasi-1D, quasi-2D and 3D materials with large localization length,
including both Coulomb repulsion and phonon-mediated attraction. The phonon-mediated attraction 
(in the pseudogapped state on the insulating side of the Superconductor-Insulator Transition) 
produces a correction to the dielectric constant, which may be detected from a linear response 
of a dielectric constant to an external magnetic field.}

\end{abstract}

\maketitle

\section{Introduction}

An Anderson insulator\cite{Anderson1958,LeeRamakrishnan1985,KramerMackinnon1993}
with a relatively long localization length $\xi$ is a paradigmatic model for 
studies of different  physical effects  due to disorder present in many poor conductors.
In particular, it was recently shown that one specific class of quantum phase transitions  between 
superconducting and insulating  states (superconductor-insulator transitions, SIT) can be understood
in terms of a relatively weak Cooper attraction between Anderson-localized electrons\cite{FIKY}. 
In such materials, superconductivity was shown to develop in two stages as temperature decreases: first, 
localized electrons are progressively bound into localized pairs 
 (forming a {\em pseudogapped insulator}\cite{FIKC}),
and then at lower temperatures macroscopic coherence between such pairs is established,
leading to superconductivity.

However, the theory developed in Refs.~\onlinecite{FIKY,FIKC} is somewhat limited 
since it neglects long-range Coulomb interaction between electron pairs (the importance of Coulomb
repulsion is known from a more standard approach to SIT based upon a model of a weak-link Josephson array
where the SIT is driven by a competition between the Josephson coupling energy and the Coulomb charging
energy\cite{SIT-B}). The first step in filling
this omission is studying the dielectric response of disordered Anderson insulators in the regime of
a large localization length (where the resulting dielectric constant may become large).
Knowledge of the dielectric constant is also important for studying high-frequency response of
superconducting resonators made of strongly disordered superconductors 
like InO$_x$, TiN and NbN, where pseudogap-related phenomena were detected\cite{sc_resonators}.

In order to specify what we understand by a ``large'' localization length, let us briefly review the
length and energy scales involved in the problem. If we denote the
electron density $n$, the average inter-electron distance is $n^{-1/d}$ (in $d$ dimensions),
and we can define the Fermi kinetic energy  (the energy scale associated with the inter-particle distance) as
\begin{equation}
\varepsilon_* = \nu^{-1} n\, ,
\label{E-Fermi}
\end{equation}
where $\nu$ is the $d$-dimensional density of states at the Fermi level.

We will work in the localized phase with the localization length $\xi$
(the decay rate of the average logarithm of the \textit{square}
of an eigenstate:  $ -\ln(|\psi(x)|^2/|\psi(0)|^2) \sim x/\xi$, assuming the maximum of the amplitude $|\psi(x)|$
at $x=0$), 
which defines the typical level spacing in the localization volume
\begin{equation}
\delta_\xi = (\nu \xi^d)^{-1}\, .
\label{delta-xi}
\end{equation}
Our large-localization-length assumption means that $\delta_\xi \ll \varepsilon_*$
(or, equivalently, the localization length is much larger than the inter-particle distance, $\xi \gg n^{-1/d}$).

In the discussion of the dielectric constant and dielectric screening, one more important scale is the
bare Coulomb interaction strength,
\begin{equation}
\varepsilon_C = e^2 n^{1/d}\, .
\label{E_C}
\end{equation}
Note that this is not the actual energy of Coulomb interactions, since this definition includes neither
dielectric screening by the localized conductance electrons (discussed in the present paper),
nor possible screening by
electrons from the valence band (which can be strong in some materials with large DoS in the valence band
close to the Fermi level).
In typical metals, $\varepsilon_C \sim \varepsilon_*$, and we may assume a similar relation here. 

Finally, in a pseudogapped insulator, one more energy scale appears: 
the pseudogap $\Delta_P$. It is assumed to be much smaller than $\delta_\xi$.

The issue of dielectric response of an Anderson insulator with  a long localization length
is not a new one, but is far from settled. Pertinent results belong to two major groups, according to the treatment
of Coulomb electron-electron interaction. The first well-known approach~\cite{Imry1,Imry2,Sadovski1,Sadovski2}
is to consider polarizability of non-interacting electrons, with an idea to add Coulomb interaction
self-consistently at the later stage.  This way one comes to the scaling dependence
for the dielectric constant
$\epsilon \propto  e^2 \nu\xi^2$, with some undefined numerical prefactor of the order of unity
(another result within the same approach belongs to Efetov~\cite{Efetov-book}, who has studied
Anderson transition at high dimensions $d \gg 1$ by means of some effective-medium theory
developed on the basis of the Bethe-lattice solution of the localization problem;
his theory predicts $\chi \propto \xi$).

An alternative approach was initiated by McMillan~\cite{McMillan} and then continued in a number of later
works ~\cite{ESB,LeeShklovski,FinReview}; 
for recent developments along similar lines see Refs.~\onlinecite{Amini,BurmistrovGornyiMirlin}.
Here, the general idea is to account for both disorder and Coulomb correlations simultaneously, as both of them
are equally important near real metal-insulator transition in 3D. Then dielectric permeability is found
to obey scaling relation $\epsilon \propto \xi^{z-1}$, where $z < 3$ is a dynamic critical exponent.
In particular, Refs.~\onlinecite{LeeShklovski,Amini} provide arguments in favor of $z \approx 2$ near 3D Anderson
transition.

We will follow the first of the approaches described above, i.e., we start from the calculation of
the polarizability $\chi$ in the model of non-interacting electrons,
with the attitude to find an accurate numerical coefficient in the $\epsilon(\xi)$ dependence.
Later on, we discuss the effects of electron-electron interactions.
The rest of the paper is organized as described below.

In Sec.~\ref{sec:polarizability}
we study the polarizability $\chi$ in the model of non-interacting electrons under the assumption
$\xi \gg n^{-1/d}$.  This problem was solved exactly in the one-dimensional case~\cite{Berezinsky,Abrikosov},
with the conclusion that $\chi \propto \xi^2$.  In higher dimensions $d>1$, no exact results are available.
Our results for $d=2,3$, based on a combination of analytical and numerical methods, confirm  
scaling $\chi \propto \xi^2$, similarly to the one-dimensional case.
As a check of the method, for $d=1$, our computations reproduce the exact result~\cite{Abrikosov}.

Then in Sec.~\ref{section:interaction}
we use our results on polarizability to study the dielectric response of interacting systems.
Here, we consider quasi-one-dimensional, quasi-two-dimensional and isotropic three-dimensional cases
and additionally require 
$\varepsilon_C \gtrsim \delta_\xi^{2/3} \varepsilon_*^{1/3}$ 
(this condition is automatically satisfied if $\xi \gg n^{-1/d}$ and $\varepsilon_C\sim \varepsilon_*$).
Under this condition, the  polarizability $\chi$ is large, thus providing an efficient dielectric screening
of Coulomb repulsion between electrons. We approximate the dielectric constant $\epsilon$
in the usual ``mean-field'' way as
\begin{equation}
\epsilon=4\pi\chi + 1\, .
\label{epsilon}
\end{equation}
This approximation neglects two potentially important effects: 
(a) mesoscopic fluctuations of dielectric screening and 
(b) interaction-induced electron-electron correlations.

The effects of type (a) may be, in principle, taken into account at the RPA level by a proper
disorder averaging. We discuss it in more detail in Section \ref{section:dielectric-constant}
and estimate that these effects should be small in quasi-1D and quasi-2D cases, but may 
become relevant in the isotropic 3D case.

The effects of type (b) go beyond RPA: they renormalize the polarization response $\chi$ itself
and are related to the Coulomb-gap effect (modification of the density of states around the Fermi
level due to interactions). These effects are considered in Section \ref{section:coulomb-gap} after
a short analysis of the relation to low-frequency conductivity in Section \ref{section:mott}.
We find that, due to dielectric screening, the Coulomb gap is renormalized to a numerically
small fraction of $\delta_\xi$, which suggests that these effects do not have large
impact on the static dielectric response.

Finally, in Section \ref{section:pseudogap}, we apply our results to the theory of pseudogapped insulators and
estimate the effect of the pseudogap on the dielectric constant $\epsilon$. Since the pseudogap suppresses
the low-frequency part of the electromagnetic response, it correspondingly reduces $\epsilon$. 
In the regime, where the pseudogap is much smaller than $\delta_\xi$, this reduction is small, but we
propose to observe it as a function of an externally applied magnetic field, where it may show up as a
linear term in the dependence of $\epsilon$  on the magnetic field.

We conclude the paper with a discussion of the results in Section \ref{section:conclusions}.

\begin{table*}[t]
\begin{tabular}{cccccccc}
Dimension & Disorder & $W$ & $\varepsilon_F$ & $\xi$ & $\nu$ & $\delta_\xi$ & $\alpha$  \\
\hline
1 & white noise & $-$ & $-$ & $-$ & $-$ & $-$ & 4.808 \\
1 & Gauss & 0.288 & 1.0 & 36.1(2)  & 0.184(5) & 0.151(5) & 4.76(9) \\
1 & Gauss & 0.288 & 1.4 & 24.7(1)  & 0.222(8) & 0.182(7) & 4.75(9) \\
1 & Gauss & 0.288 & 1.8 & 10.11(5) & 0.334(8) & 0.296(9) & 4.9(1)  \\
1 & box   & 1.3   & 0.0 & 30.6(2)  & 0.161(3) & 0.203(5) & 4.86(9) \\
1 & box   & 2.5   & 0.0 & 8.21(4)  & 0.160(4) & 0.76(2)  & 4.84(9) \\
1 & box   & 3.5   & 0.0 & 4.17(2)  & 0.159(4) & 1.51(5)  & 4.82(9) \\
\hline
2 & Gauss & 2.0   & 3.0 & 8.0(1)   & 0.084(2) & 0.185(8) & 3.3(2) \\
2 & Gauss & 2.0   & 3.5 & 5.23(6)  & 0.078(1) & 0.47(2)  & 3.2(2) \\
2 & Gauss & 2.0   & 4.0 & 3.33(3)  & 0.065(2) & 1.38(6)  & 3.2(2) \\
2 & box   & 8.0   & 0.0 & 5.50(7)  & 0.105(1) & 0.32(1)  & 3.1(2) \\
2 & box   & 9.0   & 0.0 & 3.65(4)  & 0.095(1) & 0.79(3)  & 3.2(2) \\
2 & box   & 10.0  & 0.0 & 2.66(3)  & 0.088(1) & 1.61(5)  & 3.1(2) \\
\hline
3 & Gauss & 4.0   & 7.0 & 3.35(7)  & 0.0325(3) & 0.82(6) & 2.9(2) \\
3 & Gauss & 4.0   & 7.5 & 2.05(4)  & 0.0281(2) & 4.1(2)  & 2.8(2) \\
3 & Gauss & 4.0   & 8.0 & 1.35(2)  & 0.0231(2) & 17.6(9) & 2.6(2) \\
3 & box   & 21.0  & 0.0 & 2.96(5)  & 0.0453(2) & 0.85(5) & 2.9(2) \\
3 & box   & 23.0  & 0.0 & 1.89(4)  & 0.0415(2) & 3.6(2)  & 2.8(2) \\
3 & box   & 25.0  & 0.0 & 1.44(2)  & 0.0385(2) & 8.7(4)  & 2.5(2) \\
\end{tabular}
\caption{The values of $\alpha$ calculated using Eq.~(\ref{alpha-integral}).
The model is the Anderson model on the square lattice with on-site random potential
distributed on each site independently with either Gauss or uniform-box distribution.
The disorder strength $W$ is the standard deviation for the Gauss distribution and
the width for the box distribution (see Appendix for details),
in the units of the hopping amplitude. The Fermi energy $\varepsilon_F$ is the energy
at which the wave functions are considered, in the units of the hopping amplitude,
measured from the middle of the band. The localization length $\xi$ is calculated
independently using the transfer-matrix method of Refs.~\onlinecite{McK,Slevin2014}.
The energy $\delta_\xi$ is calculated
according to Eq.~(\ref{delta-xi}). The error bars for the last digit are indicated
in brackets. The first line contains the analytical result (\ref{alpha-1D}).}
\label{table:alpha}
\end{table*}

\begin{figure}[tb]
\centerline{\includegraphics[width=0.45\textwidth]{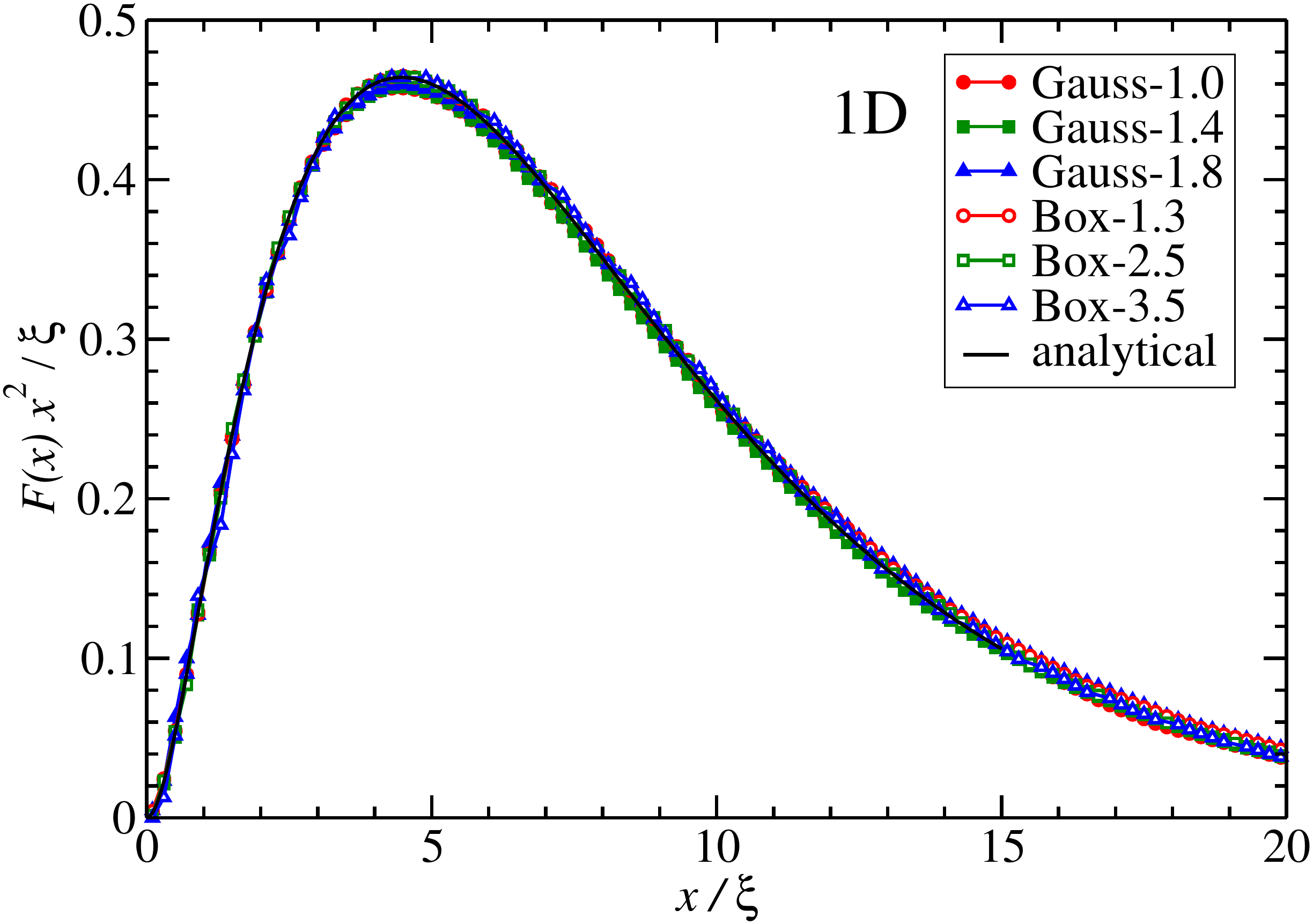}}
\centerline{\includegraphics[width=0.45\textwidth]{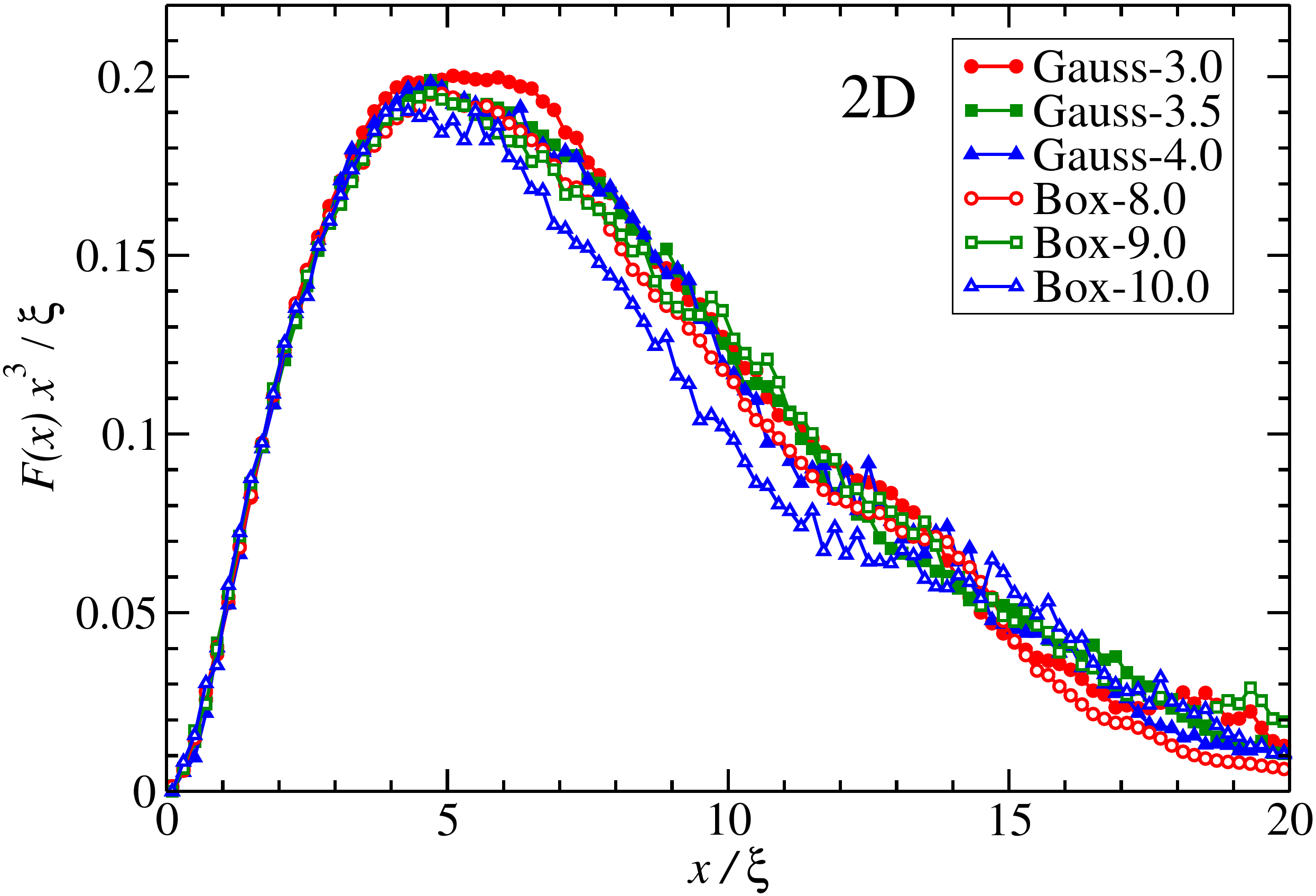}}
\centerline{\includegraphics[width=0.45\textwidth]{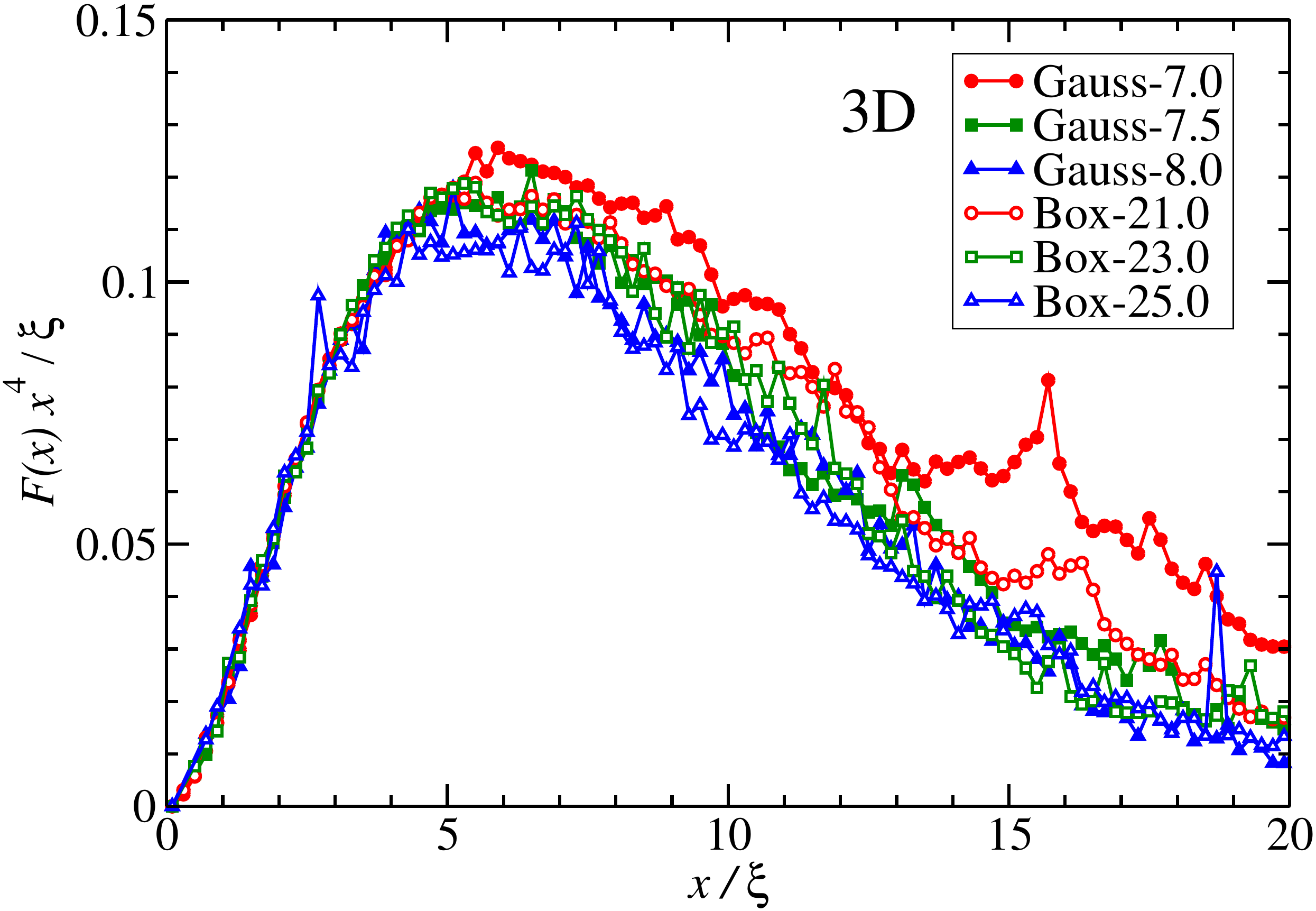}}
\caption{The function $F(x) x^{d+1}/\xi $ versus the scaled coordinate
$x/\xi$ calculated for the cases listed in Table~\ref{table:alpha}. 
The three panels show the one-, two-, and three-dimensional results, respectively.
The function $F(x)$ is multiplied by the appropriate power of $x$, in order
to reflect the contribution to the integral, Eq.($\ref{chi4}$). In order to
reduce statistical fluctuations, the points represents averages over
bins of size $0.2 x/\xi$. In one dimension, the black solid line shows
the analytical result (\ref{single-function-exact}).}
\label{fig:Fx}
\end{figure}

\section{Static polarizability of the Anderson-localized electron gas}
\label{sec:polarizability}

In this section, we show, using both analytical and numerical means, that the
static polarizability $\chi$ of disordered noninteracting fermions scales
proportionally to the square of the localization length $\xi$ in a wide range
of parameters, and compute the proportionality coefficient.

The zero-temperature polarizability of a non-interacting electron gas
in the presence of a random potential can be obtained via a perturbation theory
with respect to a uniform electric field:
\begin{multline}
\chi_{\alpha\beta}= \frac{e^2}{V} \\
\times \sum_{\varepsilon_m>\varepsilon_F \atop \varepsilon_n<\varepsilon_F} 
\frac{
\left\langle n \right| x_\alpha \left| m \right\rangle \left\langle m \right| x_\beta \left| n \right\rangle +
\left\langle n \right| x_\beta \left| m \right\rangle \left\langle m \right| x_\alpha \left| n \right\rangle
}{\varepsilon_m - \varepsilon_n}\, ,
\label{chi1}
\end{multline} 
where $\left| m \right\rangle$ and $\left| n \right\rangle $ are exact single-electron eigenfunctions with
energies $\varepsilon_{m}$ and $\varepsilon_{n}$, $\varepsilon_F$ is the Fermi level, 
$V$ is the volume of the system, $x_\alpha$ are the coordinate components, and $e$ is the electron charge.
Eq.~(\ref{chi1}) is exact, applies in any dimension, and holds for any specific realization of the random potential.
By averaging over disorder (assuming that the disorder average is isotropic) and taking the
limit  $V\to\infty$, we find
\begin{equation}
\chi= - e^2 \iint\limits_{\varepsilon_1<\varepsilon_F<\varepsilon_2}
\frac{\nu(\varepsilon_1)d\varepsilon_1 \; \nu(\varepsilon_2)d\varepsilon_2}{\varepsilon_1-\varepsilon_2}
\int d^dx \; S(\varepsilon_1, \varepsilon_2,x) \; x_1^2\, ,
\label{chi2}
\end{equation}
where the disorder-averaged response function is defined as
\begin{multline}
S(\varepsilon_1 {\ne} \varepsilon_2, x)= \frac{1}{\nu(\varepsilon_1)\nu(\varepsilon_2)} \Big\langle
\sum_{n,m} \delta(\varepsilon_n-\varepsilon_1) \; \delta(\varepsilon_m-\varepsilon_2) \\
\times \psi_n^*(x) \psi_m(x) \psi_m^*(0) \psi_n(0) \Big\rangle_{\rm dis} \, .
\label{S1}
\end{multline}
Here $\langle...\rangle_{\rm dis}$ denotes the disorder average and 
$\nu(\varepsilon)$ is the $d$-dimensional density of states (here and below we consider
{\em spinless} fermions).

Now we assume that the integrals are limited to a small region around the Fermi energy, so that the
density of states can be approximated by a constant ($\nu(\varepsilon)=\nu$) and 
$S(\varepsilon_1,\varepsilon_2, x)$ only depends on the difference $\omega=|\varepsilon_1-\varepsilon_2|$.
The characteristic energy scale for $S(\varepsilon_1, \varepsilon_2, x)$ is $\delta_\xi$,
therefore the above assumption is equivalent to $\delta_\xi \ll \varepsilon_*$,
where $\varepsilon_*$ is the characteristic energy scale of the density-of-states variations 
defined in Eq.~(\ref{E-Fermi}). This condition is satisfied for large localization lengths,
$\xi \gg n^{-1/d}$ (the localization length is larger than the inter-particle distance).

Under this assumption, we can simplify
Eq.~(\ref{chi2})  as
\begin{equation}
\chi= - e^2 \nu^2 \int_0^\infty d\omega\; \int d^dx \; S(\omega,x) \; x_1^2\, .
\label{chi3}
\end{equation}
The function $S(\omega,x)$ was studied in the one-dimensional case in Ref.~\onlinecite{GDP} and
was further conjectured in the quasi-one-dimensional case and in higher dimensions in 
Refs.~\onlinecite{Ivanov2009,Ivanov2012}. We introduce its integral over energy $F(x)$ which can
further be related to correlations of a {\em single} wave function using the orthogonality
of eigenstates:
\begin{multline}
F(x)= -2 \nu \int_0^\infty d\omega\; S(\omega, x{\ne} 0) \\
= \frac{1}{\nu} \left\langle \sum_n \delta(\varepsilon_n-\varepsilon) 
|\psi_n(x)|^2 |\psi_n(0)|^2  \right\rangle_{\rm dis} \, .
\label{S2}
\end{multline}
Thus defined quantity $F(x)$ has the dimensionality [length]$^{-d}$ and 
obeys the normalization condition
\begin{equation}
\int d^dx \; F(x) = 1\, .
\label{F0}
\end{equation}
This allows us to express $\chi$ in terms of single-wave-function statistics:
\begin{equation}
\chi= e^2 \, \frac{\nu}{2} \int d^dx \; F(x) \; x_1^2\, .
\label{chi4}
\end{equation}
The above derivation only uses the assumption that the localization length is sufficiently large
so that the polarizability is determined by a vicinity of the Fermi surface. The result should be 
valid for any type of impurities (without the single-parameter-scaling assumption) and in any dimension.

On general grounds, we expect that $F(x)$ is a function exponentially decaying at a length scale
of order $\xi$, and therefore
\begin{equation}
\chi = \alpha \, e^2 \,\nu \, \xi^2\, ,
\label{chi5}
\end{equation}
where 
\begin{equation}
\alpha = \frac{1}{2d} \int d^dx \; F(x) \; \left( \frac{x}{\xi} \right)^2
\label{alpha-integral}
\end{equation}
is a numerical coefficient depending on the dimension $d$. In principle, this coefficient
may also depend on the type of disorder and disorder strength. In the limit of white-noise
disorder, localization is described  (except in the pure 1D case)
by a non-linear supersymmetric sigma model \cite{Efetov83,Efetov-book},
and we may expect a universal form of the function $F(x)$ (more precisely, its non-oscillating part
rescaled in the units of the localization length $\xi$). 

In fact, in the one-dimensional (and
quasi-one-dimensional) case, the function $F(x)$ in the white-noise limit is known analytically 
\cite{Gogolin1976,LGP82,Kolokolov1995,Efetov-book,Mirlin1997,EL1983,Ivanov2009}:
\begin{equation}
F(x)=
  2\pi^2 \xi \; \frac{\partial^2}{\partial x^2}
  \int_0^\infty k \, dk \, \frac{\tanh\pi k}{\cosh^2\pi k}
  e^{-(k^2+1/4)x/\xi}\, .
\label{single-function-exact}
\end{equation}
On substituting (\ref{single-function-exact}) into (\ref{alpha-integral}),
 we integrate by parts twice to obtain
\begin{equation}
\alpha=4\pi^2 \int_0^\infty dk \, \frac{k}{k^2+1/4} \, \frac{\tanh\pi k}{\cosh^2\pi k}
= 4\zeta(3)=4.808\ldots
\label{alpha-1D}
\end{equation}
Abrikosov and Ryzhkin in Ref.~\onlinecite{Abrikosov} present an equivalent formula [below their Eq.~(2.95)];
an apparent difference by a factor of 2
from our result is due to the spin degeneracy accounted for in Ref.~\onlinecite{Abrikosov}.

\begin{figure}[tb]
\centerline{\includegraphics[width=0.45\textwidth]{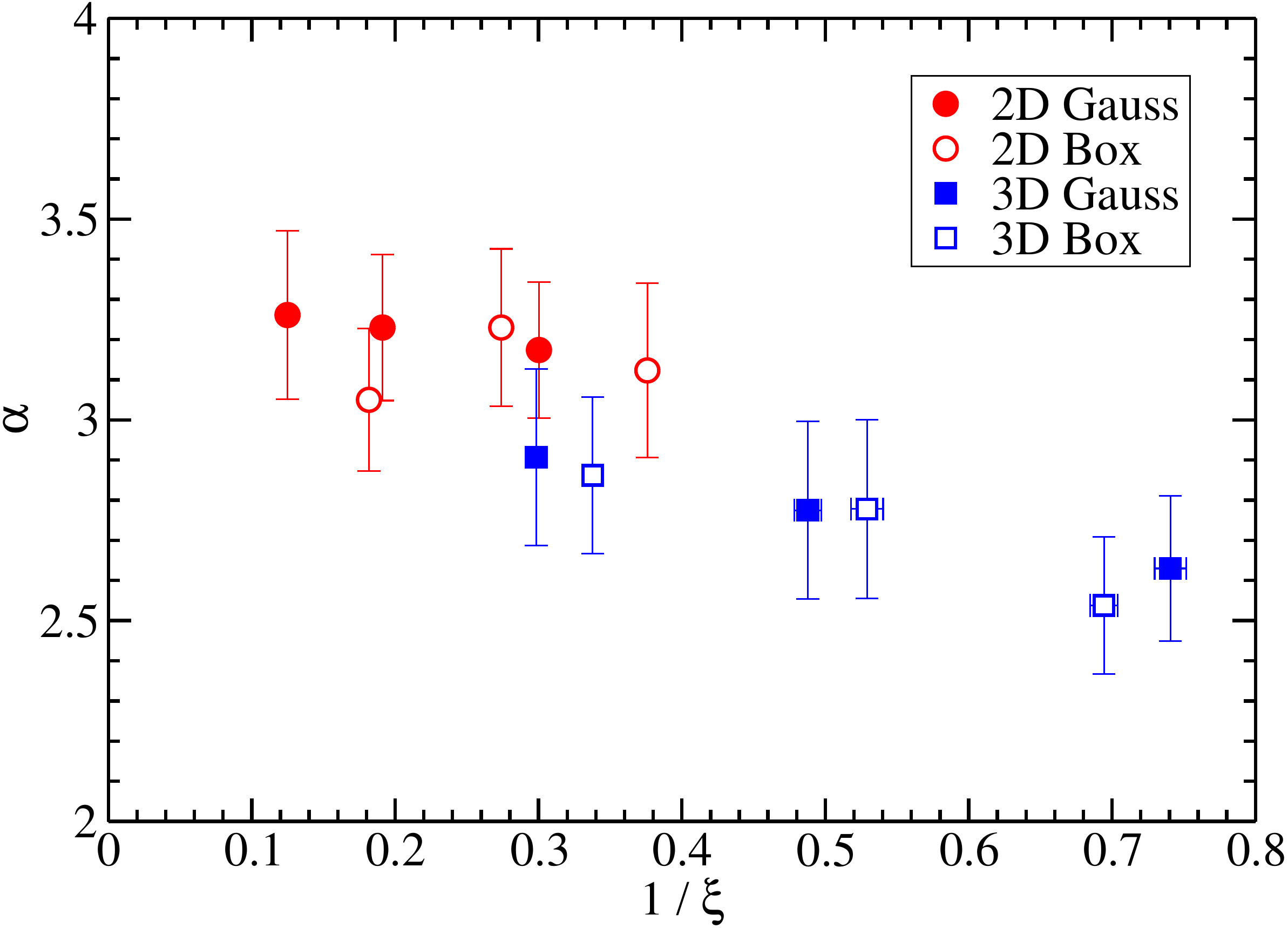}}
\caption{Values of the proportionality coefficient $\alpha$ in Eq.~(\ref{chi5})
in two and three dimensions as a function of the inverse localization length.}
\label{fig:alpha}
\end{figure}

We have further studied numerically the function $F(x)$ and the coefficient $\alpha$ for two models
of disorder at different strengths of disorder levels in one, two, and three dimensions. Namely,
we considered the so-called Anderson model: the tight-binding model on the square lattice (linear chain
or cubic lattice, depending on dimension) with a random on-site disorder (see Appendix for details).
The distribution of disorder
was taken either uniform-box or Gaussian. We have analyzed a variety of levels of disorder strength at
different levels of $\varepsilon_F$ and have found that $F(x)$ is nearly universal when rescaled in
length units of $\xi$. 
Our results for $F(x)$ (multiplied by a suitable power of $x$) are shown in Fig.~\ref{fig:Fx} 
and the corresponding values of $\alpha$ are reported in Table~\ref{table:alpha} and
in Fig.~\ref{fig:alpha}.

In one dimension, the numerical results for $\alpha$ agree with the analytical value (approximately $4.8$).
In two and three dimensions, the value of $\alpha$ is nearly universal ($3.2$ in two dimensions
and $2.8$ in three dimensions) for the studied range of parameters. A slight trend to the increase 
of $\alpha$  with increasing $\xi$ in 3D may be either real or an artifact of our numerical methods.
In the former case, the extrapolated value of $\alpha$ at the localization transition $\xi\to\infty$
is about $3.1$. 

Note that $F(x)$ has qualitatively very similar behavior in all the three dimensionalities. In all the three
cases, the main contribution to the integral (\ref{chi4}) comes from large distances, of the order
of 5--10 localization lengths. At such large distances, the function $F(x)$ is determined not by typical
localized states, but by  relatively rare Mott-hybridization events \cite{Ivanov2012}. As a consequence, $F(x)$
decays not at the length scale $\xi$, but at a larger one (at the length scale of $4\xi$ in one dimension
in the white-noise limit; we are not aware of similar exact results in higher dimensions).

\section{Effects of interaction} 
\label{section:interaction}

\subsection{Dielectric constant}
\label{section:dielectric-constant}

The calculation of the previous section was for a single-particle response to an external field.
It only assumed that the localization length is much larger than the inter-particle spacing,
$\xi \gg n^{-1/d}$.  In a physical material, one is instead interested in the dielectric
constant $\epsilon$ describing the dielectric screening of the external field by polarized localized
states and reducing the Coulomb interaction between electrons. In three dimensions, the
result (\ref{chi5}) may be rewritten as
\begin{equation}
\chi=\alpha \, \varepsilon_C \, \delta_\xi^{-2/3} \, \varepsilon_*^{-1/3}\, ,
\end{equation}
using the energy scales (\ref{E-Fermi})--(\ref{E_C}) defined in the Introduction. Under
the assumption $\delta_\xi \ll \varepsilon_* \sim \varepsilon_C$, we find $\chi\gg 1$,
and then the dielectric constant (\ref{epsilon}) is  even larger by an extra order of magnitude.

However, 
the standard formula (\ref{epsilon}) implies the uniformity of the \textit{actual} 
electric field inside in the sample, which is not obviously granted near Anderson transition.
This problem exists already within the standard RPA approach: the screened Coulomb potential 
$V(\mathbf{r},\mathbf{r'})$ for a given realization of disorder is determined (within the RPA) by the equation
\begin{equation}
( 4\pi e^2)^{-1} \nabla^2 V(\mathbf{r},\mathbf{r'}) + \int d^2r_1 \Pi(\mathbf{r},\mathbf{r}_1) 
V(\mathbf{r}_1,\mathbf{r'})  = \delta(\mathbf{r}-\mathbf{r'})
\label{1}
\end{equation}
where $\Pi(\mathbf{r},\mathbf{r}_1) = \lim_{\omega\to 0} \Pi(\mathbf{r},\mathbf{r}_1;\omega) $ 
is the polarization  function of the free Fermi gas with random potential. 
Then, the dielectric constant $\epsilon$ is determined by the long-scale  asymptotics of the averaged potential 
\begin{equation}
\overline{V(\mathbf{r},\mathbf{r'})} \approx
\frac{e^2}{\epsilon |\mathbf{r}- \mathbf{r'}| }
\label{Vef}
\end{equation}
The procedure we used in Sec.~\ref{sec:polarizability} to calculate  the polarizability $\chi$
was equivalent to averaging the polarization function,  $\Pi(\mathbf{r},\mathbf{r}_1;\omega) \to
 \overline{\Pi(|\mathbf{r}-\mathbf{r}_1|;\omega)}$, instead of  averaging the resulting potential
$V(\mathbf{r},\mathbf{r'})$.  Physically, the problem with such a preliminary averaging is the it neglects
the fact that the \textit{microscopic} electric field is inhomogeneous
in a disordered sample.  This effect is expected to be important near  3D  Anderson transition, 
due to strong fractality of electron wave functions with energies near the mobility
edge, see Ref.~\onlinecite{Kravtsov2007} and references therein. 

In low-dimensional systems with long localization length,
wave-function fractality  is absent ($d=1$), or rather weak ($d=2$). Thus averaging over disorder,
already  at the level of the polarization function, should not lead to any considerable error. 
Therefore our results (\ref{chi5}), (\ref{alpha-integral}) imply direct consequences for the macroscopic
dielectric constant in quasi-1D  materials (conducting 1D wires with moderate disorder in a poorly conducting matrix)
and quasi-2D (layered) materials with relatively large Drude sheet conductance $g \geq 1$ and very poor interlayer 
tunneling.  In both these cases,  the material becomes insulating at low temperature, and
we provide  expressions for bulk dielectric constants, assuming the direction of  the electric field
 to be parallel to conducting  wires or planes:
\begin{eqnarray}
\label{eps1}
 \epsilon_1 =  16 \pi \zeta(3) e^2 \nu_1 n_{2} \xi^2   \qquad  (\mathrm{quasi-1D})  \\
 \epsilon_2 =  4\pi \alpha_2 e^2 \frac{\nu_2} {d} \xi^2 \qquad (\mathrm{quasi-2D})
\label{eps2}
\end{eqnarray}
where  $n_2$ is the two-dimensional density of wires in the quasi-1D sample, 
$\alpha_2 \approx 3.2$, and $d$ is the interlayer distance in a quasi-2D material
(note that Eq.~(\ref{eps1}) is known since Ref.~\onlinecite{Abrikosov}).
Note that the materials we discuss here are considered to be  quasi-1D or quasi-2D with regard to
their electronic properties only; in particular,  typical  interchain (interplane) length scales 
$(n_2^{-1/2}, d)$ are only slightly larger than the interatomic distances along 
highly conductive directions.
Therefore, large values of 1D or 2D polarizability  also imply large values of $\epsilon_{1,2}$ 
as presented in Eqs.\ (\ref{eps1}), (\ref{eps2}).

In a bulk (isotropic 3D) system close to localization threshold, the expression
\begin{equation}
\epsilon = 4\pi \alpha_3 e^2 \nu \xi^2  \qquad \alpha_3 \approx 3
\label{eps3}
\end{equation}
may serve just as an order-of-magnitude estimate until a solution  to the 
random RPA equation (\ref{1}) with a more accurate disorder averaging is available.
We will further use Eq.~(\ref{eps3}) to estimate possible effects of Coulomb interaction
which go \textit{beyond RPA}.

\subsection{Low-frequency conductivity and Kramers-Kronig relation}
\label{section:mott}

Effects of interactions between electrons with the characteristic energy scale
below $\delta_\xi$ may be estimated from the frequency-dependent conductivity via
the Kramers-Kronig relation. At frequencies below $\delta_\xi$, 
the conductivity is expected to obey the Mott-Berezinsky law,
\begin{equation}
\sigma(\omega) \approx \Sigma_d \, \frac{e^2}{\hbar} \, \xi^{2-d} \, \left(\frac{\omega}{\delta_{\xi}}\right)^2
\ln^{d+1}\frac{\tilde\delta_\xi}{\omega}\, ,
\label{MB-conductivity}
\end{equation}
where $\tilde{\delta}_{\xi} = c_d\delta_\xi$ and $c_d$ is some number of the order of unity.
In the 1D case both numerical coefficients  are known: $\Sigma_1 = 4\pi$ (for spinless fermions),
and $c_1 = 1/\pi$, see Ref.~\onlinecite{Abrikosov}.

Now we can use the Kramers-Kronig relation
\begin{equation}
\chi=\frac{2}{\pi}\int_0^\infty \frac{d\omega}{\omega^2} \sigma(\omega)\, ,
\label{KK-relation}
\end{equation}
in order to estimate the dominant contribution to $\chi$ and compare it with our results obtained
in the previous Section. Namely, we approximate Eq.(\ref{KK-relation}) as
\begin{equation}
\chi \approx \frac{2}{\pi}\int_0^{\tilde{\delta}_\xi} \frac{d\omega}{\omega^2} \sigma(\omega)
 \approx \frac{2(d+1)!}{\pi} \, \Sigma_d \, c_d \, e^2 \, \nu \, \xi^2
\, ,
\label{KK-2}
\end{equation}
assuming that the dominant contribution to $\chi$ comes from resonant Mott pairs with 
$\omega < \tilde{\delta}_\xi$, see Eq.~(\ref{MB-conductivity}).
 This  is in agreement 
with the shape of the function $F(x)$ shown in Fig.~\ref{fig:Fx}, where the region of
 $ x \sim 5-10 \, \xi$ provides the  main contribution to the  integral (\ref{alpha-integral}).

Using the Kramers-Kronig relation (\ref{KK-2}) and our results
for $\chi$ in Sec.~\ref{sec:polarizability}, we can relate the 
coefficients $\Sigma_d$ and $c_d$ in Eq.~(\ref{MB-conductivity}) (for spinless fermions) as
\begin{equation}
\alpha_d \approx \frac{2(d+1)!}{\pi} \, \Sigma_d \, c_d\, ,
\label{Sigma-d}
\end{equation}
up to a numerical coefficient of order one. For $d=1$ the above relation can be 
tested by means of exact results: it  is accurate within a  6\% error.

\subsection{Effects of Coulomb repulsion}
\label{section:coulomb-gap}

Our calculations of $\chi$ do not take into account Coulomb interactions between localized electrons.
Such interactions, due to their long-range nature, modify the density of states close to the
Fermi energy (the so called Coulomb gap) and the low-frequency dynamics.
In three dimensions, the magnitude of the Coulomb gap is estimated as~\cite{ES1975}
\begin{equation}
\Delta_{C} \approx \frac{e^3 \nu^{1/2}}{\epsilon^{3/2}}\, .
\label{charge-gap}
\end{equation}
In our case, the large dielectric constant  $\epsilon$ is produced by the same localized
electrons, and, using (\ref{eps3}), we estimate
\begin{equation}
\Delta_{C} \approx \frac{\delta_\xi}{(4\pi\alpha)^{3/2}} \approx \frac{\delta_\xi}{250}\, .
\label{charge-gap-comparison}
\end{equation}

In other words, the Coulomb-gap scale is only by a numerical factor smaller than the localization
energy scale $\delta_\xi$. Our estimate of this smallness (\ref{charge-gap-comparison}) should
be taken with care, since both $\delta_\xi$ and the Coulomb gap are only estimated up to a numerical
coefficient of order one. If it turns out that $\Delta_{C}$ is of the same magnitude as $\delta_\xi$,
our calculations would need to be revised by self-consistently taking into account
the effect of Mott-hybridized states on the electrostatic potential.

An attempt to take into account the effect of the Coulomb repulsion on the Mott conductivity
(\ref{MB-conductivity}) was done in Ref.~\onlinecite{ES1981}. Even though their results
are sometimes used in the discussion of  conductivity in general-type Anderson insulators
(or electron glasses)~\cite{Gruner2004}, they were originally derived for doped semiconductors with
strongly localized electronic states. As we show below, in our case (large $\xi$ and, consequently,
large dielectric constant $\epsilon$, see Section \ref{section:dielectric-constant}),
the applicability regime of the results of Ref.~\onlinecite{ES1981} is small or vanishes.

Indeed, according to Ref.~\onlinecite{ES1981}, the Coulomb interaction between electrons should
be taken into account by replacing one of the $\omega$ factors in Eq.~(\ref{MB-conductivity})
by $\omega + e^2/ r_\omega \epsilon$, where $r_\omega = 2\xi\ln(\tilde\delta_\xi / \omega)$ 
is the Mott hybridization length. At low enough $\omega$ the additional term $e^2/ r_\omega \epsilon$
becomes dominant, which changes frequency-dependence of conductivity crucially, in comparison with 
Eq.~(\ref{MB-conductivity}), $\sigma(\omega) \propto \omega \ln^d(\tilde\delta_\xi/\omega)$.
Clearly, this dependence cannot be valid down to arbitrary low $\omega$,  otherwise
the integral (\ref{KK-relation}) would diverge. Indeed it is explained in Ref.~\onlinecite{ES1981}
that applicability of this result is limited by the condition
\begin{equation}
 \frac{e^2}{\epsilon r_\omega} \gg \Delta_{C}\, .
\label{conditionES}
\end{equation}
otherwise the Coulomb-gap effect needs to be solved self-consistently.

Thus the result of Ref.~\onlinecite{ES1981} contains two opposite conditions on the relevant frequencies.
The first limitation is $e^2/ r_\omega \epsilon \ge \omega$ (otherwise the effect of the Coulomb correction is small).
This condition translates, in our case of large $\xi$, into
\begin{equation}
\omega \le \frac{\delta_\xi}{8\pi\alpha} \approx 0.01\delta_\xi\, .
\end{equation}
The other condition is the actual applicability limitation (\ref{conditionES}), which translates into
\begin{equation}
\omega \gg  c_3 \delta_\xi\exp\left(-\sqrt{\pi\alpha}\right) \approx 0.05 c_3 \delta_\xi\, .
\label{low-frequency-limitation}
\end{equation}
As we see, those conditions are both formulated in terms of a numerically small fraction of $\delta_\xi$
of similar magnitude. Therefore, in our situation of large $\xi$, the range of applicability of the results
of Ref.~\onlinecite{ES1981} practically vanishes (the above discussion has no bearing on the original
application of results of Ref.~\onlinecite{ES1981} to the strongly localized regime with short $\xi$).

Note that the condition (\ref{low-frequency-limitation}) also sets the limit for
the applicability of the Mott-Berezinsky formula (\ref{MB-conductivity}).

\subsection{Effects of phonon-mediated attraction: Anderson insulators with a pseudogap}
\label{section:pseudogap}

Based on the 
Kramers-Kronig relation, we can analyze the dielectric
response of an Anderson insulator with a local phonon-mediated 
attraction~\cite{Ma1985,Trivedi,FIKY,FIKC}. Such an attraction leads to a ``pseudogap''
in the single-particle density of states~\cite{Sacepe2011}
(similar to the ``parity gap'' in ultra-small metallic grains~\cite{MatveevLarkin1997}).
The energy scale of the pseudogap $\Delta_P$ is given by the typical interaction energy of
a localized state. We assume that $\Delta_P \ll \delta_\xi$, but not too small, in order
to neglect the effects of Coulomb interaction [in three dimensions, we estimate
this condition to be $\Delta_P \gg 0.05 \tilde\delta_\xi$ from Eq.~(\ref{low-frequency-limitation})].

The influence of the pseudogap upon frequency-dependent Mott conductivity of an insulator was studied
by two of the present authors in Ref.~\onlinecite{IF2017}. It was found  that asymptotic form
(\ref{MB-conductivity}) remains valid also below $\Delta_P$, but with a strongly
reduced prefactor $\Sigma_d$ (by a factor more than 10). We can therefore estimate the
effect of the pseudogap on the dielectric susceptibility (proportional to $\chi$) by simply
excluding the low-frequency part $\omega<\Delta_P$ from the Kramers-Kronig integral (\ref{KK-relation}).

This leads to the following result for the reduction of the dielectric response due to the pseudogap:
\begin{equation}
 -\frac{d\ln\epsilon}{d\Delta_P} \sim \frac{8}{\epsilon} \frac{\sigma(\omega)}{\omega^2}\Big|_{\omega=\Delta_P}
\sim \frac{1}{\delta_\xi} \left(\ln\frac{\delta_\xi}{\Delta_P}\right)^{d+1}
\label{dedD}
\end{equation}
(up to a numerical coefficient of order one).

This effect of the pseudogap on the dielectric response may be convenient to observe by applying
an external magnetic field. A magnetic field $B$ suppresses the parity gap via Zeeman effect, 
\begin{equation}
\Delta_P(B) = \Delta_P - g_{\mathrm{eff}}\,\mu_B\, B\, .
\label{mu}
\end{equation}
The effective Lande factor $g_{\mathrm{eff}}$ is material-dependent: it may vary from its
bare value 2 in the absence of spin-orbit coupling to small values of order  $\tau_{\mathrm so}\Delta_P/\hbar \ll 1$, 
if strong spin-orbit scattering is present
($\tau_{\mathrm so}$ is the spin-order scattering time)~\cite{Anderson1959,AG1962}.
Combining Eqs.\ (\ref{dedD}) and (\ref{mu}) we find a \textit{positive linear} effect of small magnetic fields 
upon $\epsilon$:
\begin{equation}
 \frac{d\ln\epsilon}{d B} \sim 
 \frac{g_{\mathrm{eff}}\,\mu_B}{\delta_\xi} \left(\ln\frac{\delta_\xi}{\Delta_P(B)}\right)^{d+1}
\label{dedB}
\end{equation}

Another effect of the magnetic field on the dielectric constant $\epsilon$ is
via modification of the localization length $\xi$ due to orbital effect. 
However, the corresponding  $\xi(B)$ dependence is quadratic at
weak magnetic fields, $\delta\xi(B)/\xi \sim (B\xi^2/\Phi_0)^2$, therefore the pseudogap-related
contribution (\ref{dedB}) dominates at low fields.

The results (\ref{dedD}) and (\ref{dedB}) should also be applicable for two-dimensional films, but the
applicability condition at low $\Delta_P$  [replacing Eq.~(\ref{low-frequency-limitation})]
requires further study.

\section{Conclusions and discussion}
\label{section:conclusions}

The main result of the paper is Eq.~(\ref{chi5}) for the polarizability of a non-interacting
 Anderson insulator with a large localization length. This result was derived by a combination
of analytic and numerical methods in dimensionalities $d=1,2,3$, and the
coefficient $\alpha$ was computed in all the three cases (in one dimension,
our computation agrees with the exact analytical result).
The main contribution to polarizability comes from
 relatively large distances (of order 5-10 localization lengths),
which implies an important contribution from relatively rare
Mott-hybridized eigenstates \cite{Ivanov2012}.
As a by-product of this study, we found that the rescaled correlation function 
$F(x) x^{d+1}$ is of  nearly universal shape in all dimensions $d=1,2,3$, 
see Fig.~\ref{fig:Fx}.

It follows from our main result that, in case of a large localization length, 
the polarizability is also large.  For  quasi-1D and quasi-2D materials
it means also very large dielectric constant, see Eqs.~(\ref{eps1}), (\ref{eps2}).
Therefore Coulomb interactions between electrons in these materials  are strongly reduced, 
and all the energy scales arising from Coulomb interactions are renormalized down to a 
numerically small fraction of $\delta_\xi$.
Note however that the actual relevant energy scale is also a certain fraction of
$\delta_\xi$ (we denote it $\tilde\delta_\xi$ in Section \ref{section:mott}). Thus
the role of Coulomb effects may be larger than expected. However, since all the
relevant energy scales are numerical fractions of $\delta_\xi$, we expect that
the renormalization of the dielectric response due to Coulomb-gap effects
is also only by a numerical factor of order one, and therefore our results (\ref{eps1}),
(\ref{eps2}) give correct order-of-magnitude estimates.

For isotropic 3D materials with large but finite localization length, 
similar conclusions  should be taken with care,  as explained in 
Sec.~\ref{section:dielectric-constant}.   Indeed, the numerical results\cite{Amini}
obtained within Hartree-Fock scheme indicate a crucial role of Coulomb 
interaction in the pinning of the mobility edge to the Fermi-level in 
such a system. Similar conclusions about the importance of Coulomb interactions
come out of the Renormalization-Group study~\cite{BurmistrovGornyiMirlin}. 
If such a situation -- with pinned mobility edge -- is realized, then Coulomb interaction is
strongly relevant while our approach is not adequate. 
One should note, however, that the value of $z$ for the 3D metal-insulator transition with Coulomb  
interaction is still  unknown. Although numerical data~\cite{Amini} suggest
$z \approx 2$, it is not evident that the size of the system used in these computations is
sufficient to make any definite conclusion, since, according to our results (see Fig. 1), 
the major contribution to polarizability comes from large spatial scales $r \sim 5-10 \,\xi$.

In some disordered materials a large dielectric constant 
$\epsilon \gg 1$
is already provided by the contribution of valence-band electrons with energies separated from the Fermi-level
by some nonzero but relatively small gap  $E_0 \ll E_F$. Then Coulomb interaction between electrons
in the conduction band (subject to Anderson localization due to strong disorder) is
reduced accordingly.  In such a case, two different scenarios may be realized: 
a)  unpinned mobility edge with "non-interacting scaling" \,  $\epsilon \propto \xi^2$, 
that is realized at moderately large $\xi \leq \xi_1$ (for some
crossover scale $\xi_1$),
and b)  at further increase of
$\xi$ beyond $\xi_1$, a pinning of the mobility edge to the Fermi level occurs,
which is accompanied by a crossover to the "interacting scaling" similar to the one described
in Refs.~\onlinecite{Amini,BurmistrovGornyiMirlin}.

We  also note the difference between our case and the findings
of Ref.~\onlinecite{ES1981} (derived in the regime of small localization lengths)
where Coulomb interactions modify the frequency-dependent conductivity drastically
in a wide range of frequencies.

One of the applications of our results is for Anderson insulators with a 
phonon-mediated pseudogap. In such materials, the dielectric constant is
slightly reduced due to the pseudogap and, together with the pseudogap, may be tuned by an
external magnetic field. We predict a linear dependence of this correction
on the magnetic field and estimate the magnitude of this effect.

The major remaining challenge in these topics is the simultaneous account for
wave-function fractality and Coulomb-gap effects.  This is a many-body problem,
which goes beyond the methods of the present paper.

We are grateful to I.~S.~Burmistrov and V.~E.~Kravtsov for useful discussions
and to A.~D.~Mirlin, M.~V.~Sadovskii and B.~I.~Shklovskii for their comments on the 
first version of the manuscript.
This research was supported by the Murcia Regional
Agency of Science and Technology (project 19907/GERM/15).
The work of D.A.I.\ was partly supported by the
Swiss National Foundation through the NCCR QSIT.
The work of M.V.F.\ was supported by the Russian
Science Foundation grant \mbox{14-42-00044}.

\appendix

\section{Details of numerical calculations}
\label{appendix-1}

For numerical calculations of Section~\ref{sec:polarizability}, 
we sampled single-particle eigenstates in the tight-binding
model on the cubic lattice (in $d$ dimensions) with the random on-site potential.
The Hamiltonian has the usual form:
\begin{equation}
H=-\sum_{\langle i,j\rangle}a^\dagger_i a_j + \sum_i V_i a^\dagger_i a_i\, ,
\end{equation}
where the first term describes nearest-neighbor hopping and the second term the random potential
($a^\dagger_i$ and $a_i$ are the creation and annihilation operators). The hopping amplitude
is taken to be one. The on-site disorder $V_i$ is uncorrelated on different sites and is drawn
from the normal distribution centered at zero with the standard deviation $W$ for ``Gaussian''
models and from the uniform distribution between $-W/2$ and $W/2$ for ``Box'' models.

For calculating $F(x)$ and $\alpha$, we used systems of cubic geometry with periodic boundary
conditions. The linear system sizes were equal to 2000 for one-dimensional case,
750 for two dimensions and 100 for three dimensions. The data for each system were obtained by
averaging over about 1500 eigenstates in many disorder realizations  in 2D and 3D and over
$3.2\times 10^6$ eigenstates in 1D.

The localization length $\xi$ was computed independently using the transfer-matrix
method of Refs.~\onlinecite{McK,Slevin2014}. Note that  the definition of the localization
length usually employed while using this method is two times larger than the one 
we used here.

\end{document}